\documentclass[aps,pre,notitlepage,amsmath,amssymb]{revtex4-1}
\usepackage{graphicx}

\begin{document}

\title{Stochastic dynamics of $N$ correlated binary variables and non-extensive
statistical mechanics}

\author{A. Kononovicius}
\email{aleksejus.kononovicius@tfai.vu.lt}
\affiliation{Institute of Theoretical Physics and Astronomy, Vilnius University,
A. Go\v{s}tauto 12, LT-01108 Vilnius, Lithuania}

\author{J. Ruseckas}
\affiliation{Institute of Theoretical Physics and Astronomy, Vilnius University,
A. Go\v{s}tauto 12, LT-01108 Vilnius, Lithuania}

\date{\today{}}

\begin{abstract}
The non-extensive statistical mechanics has been applied to describe a variety
of complex systems with inherent correlations and feedback loops. Here we
present a dynamical model based on previously proposed static model exhibiting
in the thermodynamic limit the extensivity of the Tsallis entropy with $q<1$ as
well as a $q$-Gaussian distribution. The dynamical model consists of a
one-dimensional ring of particles characterized by correlated binary random
variables, which are allowed to flip according to a simple random walk rule. The
proposed dynamical model provides an insight how a mesoscopic dynamics
characterized by the non-extensive statistical mechanics could emerge from a
microscopic description of the system.
\end{abstract}
\maketitle

\section{Introduction}

Assumption that the velocities of the colliding particles are uncorrelated and
independent of the position of the particles is a key point in the molecular
chaos hypothesis. The hypothesis itself is a cornerstone of the classical
(extensive) statistical mechanics. In a more general form the lack of
correlations assumption is included into the functional form of the
Boltzmann-Gibbs entropy: 
\begin{equation}
S_{\mathrm{BG}}=-\int P(x)\ln P(x)\,dx\,.\label{eq:BG}
\end{equation}
This assumption works rather well for many classical dynamical systems, in which
large number of particles and their interactions helps to mask the existing
correlations. E.g., after the collision of gas particles the velocities of those
particles are no longer uncorrelated, but due to large number of them and, thus,
large number of the collisions between them, the correlation quickly becomes
forgotten. Yet there are systems in which long-range interactions, long-range
memory or non-ergodicity are present. To understand these, sometimes referred to
as complex, systems and their apparently anomalous properties a generalization
of classical statistical mechanics proposed by Tsallis \cite{Tsallis2009-1} is
used. This generalized framework has found its applications to extremely
different systems studied by both ``hard'' (e.g., mathematics, physics,
chemistry) and ``soft'' (e.g., economics) sciences
\cite{Gell-Mann2004,Adare2011,Afsar2013,Lutz2013,Vallianatos2013,Gontis2014,Tanaka2015,Jauregui2015}.

The non-extensive statistical mechanics framework is constructed starting from
the generalization of the Boltzmann-Gibbs entropy\cite{Tsallis2009-1}
\begin{equation}
S_{q}=\frac{1-\int[P(x)]^{q}\,dx}{q-1}\,,\label{eq:q-entr}
\end{equation}
where $P(x)$ is a probability density function of finding the system in the
state characterized by the parameter $x$, the parameter $q$ describes the
non-extensiveness of the system. The original Boltzmann-Gibbs entropy,
Eq.~(\ref{eq:BG}), can be obtained from Eq.~(\ref{eq:q-entr}) in the limit
$q\rightarrow1$ \cite{Tsallis2009-1,Tsallis2009-2}. More generalized entropies
and distribution functions are introduced in
Refs.~\cite{Hanel2011-1,Hanel2011-2}. In the classical statistical mechanics the
Gaussian distribution plays an important role. Similarly, in the non-extensive
statistical mechanics the $q$-Gaussian distribution
\begin{equation}
P_{q}(x)=C\exp_{q}(-A_{q}x^{2})\label{eq:q-gauss}
\end{equation}
becomes important \cite{Tsallis2009-1}. Note that in Eq.~(\ref{eq:q-gauss})
$q$-exponential function is present, this function is defined as follows
\begin{equation}
\exp_{q}(x)\equiv[1+(1-q)x]_{+}^{\frac{1}{1-q}}\,,\label{eq:q-exp1}
\end{equation}
where the notation $[x]_{+}$ means $[x]_{+}=x$ if $x>0$, and $[x]_{+}=0$
otherwise. Recently, in Ref.~\cite{Ruiz2015} it has been shown that $q$-Gaussian
distributions emerge in the limit of large number of realizations in a
generalized binomial distribution representing a sequence of correlated trials.

To illustrate non-extensive statistical mechanics several statistical models
which in the $N\rightarrow\infty$ limit provide $q$-Gaussian attractors have
been constructed \cite{Rodriguez2008,Hanel2009,Rodriguez2012,Ruseckas2015}. The
first attempt to create such models has been undertaken in
Ref.~\cite{Moyano2006}. However, it has been shown that the distributions do not
approach a $q$-Gaussian form when the number of particles $N$ in the model
increases \cite{Hilhorst2007}. Two models that provide $q$-Gaussian
distributions have been introduced in \cite{Rodriguez2008}, the second model
does so by construction. More detailed analysis of the models in
\cite{Rodriguez2008} have been presented in \cite{Hanel2009} and the
generalization to higher dimensions has been proposed in \cite{Rodriguez2012}.
However, the standard Boltzmann-Gibbs entropy remains extensive for the models
from Ref.~\cite{Rodriguez2008}. This situation has been improved by the model
presented in Ref.~\cite{Ruseckas2015}. This model is based on a system composed
of $N$ distinguishable particles arranged in a chain, each of the particles in
the chain is characterized by a binary random variable (this can be, for
example, particles with the spin $\frac{1}{2}$). In the aforementioned model it
is assumed that spins next to each other are almost always aligned in the same
direction. The only exceptions are assumed to be $d$ cases in which the
neighboring spins are antialigned. In this model the number of states grows as a
power-law, with the corresponding index $q$ being $q_{\mathrm{stat}}=1-1/d$. In
addition, the distribution of the total spin of the system in the limit of
$N\rightarrow\infty$ tends to a $q$-Gaussian with $q_{\mathrm{dist}}\neq
q_{\mathrm{stat}}$. All models in \cite{Moyano2006,Rodriguez2008,Ruseckas2015},
except the last model of \cite{Rodriguez2008} are for $q\leq1$.

The distributions of the non-extensive statistical mechanics can be obtained
from mesoscopic description of the systems in the form of probabilistic dynamics
\cite{Tsallis2009-1}. Such probabilistic description can be provided by
nonlinear Fokker-Planck equations and corresponding nonlinear stochastic
differential equations (SDEs) \cite{Borland1998,Borland2002}, SDEs with additive
and multiplicative noises \cite{Anteneodo2003,Santos2010} or with multiplicative
noise only \cite{Queiros2007}, and with fluctuating friction forces
\cite{Beck2001}. Nonlinear SDEs generating distributions of non-extensive
statistical mechanics together with $1/f$ noise have been proposed in
Ref.~\cite{Ruseckas2011}. Scaling law that follows from nonlinear Fokker-Planck
equations has been recently experimentally confirmed in granular media
\cite{Combe2015}.

The goal of this paper is to extend the model presented in
Ref.~\cite{Ruseckas2015} by introducing stochastic temporal dynamics. Conserving
the total number of domains $d$, we allow spins near domain boundaries to flip.
This effectively introduces a random walk of domain boundaries into the previous
static model. In the limit of large number of spins we derive the Fokker-Planck
equation and a corresponding SDE describing the temporal dynamics of the total
spin of the system. The proposed dynamical model provides an insight how a
mesoscopic dynamics characterized by the non-extensive statistical mechanics
could emerge from a microscopic description of the system.

The Letter is organized as follows. To show how stochastic temporal dynamics can
be introduced into a model describing the equilibrium, in
Section~\ref{sec:uncorrelated} we introduce stochastic dynamics into a simple
model consisting of uncorrelated binary random variables. In
Section~\ref{sec:correlated} we use a similar stochastic temporal dynamics for
the model with correlated binary random variables and leading to extensive
generalized entropy with $q<1$. In Section~\ref{sec:long-range} we demonstrate
that presence of macroscopic fluctuations in this model can lead to $1/f$ noise,
whereas in Section~\ref{sec:subsyst} we consider properties of a part of larger
system. We summarize the paper in Section~\ref{sec:conclusions}.

\section{Stochastic dynamics in the model with uncorrelated binary variables}

\label{sec:uncorrelated}As in Ref.~\cite{Ruseckas2015}, we start with the
analysis of the dynamical model without any correlations between the particles
or their spins. Let us say that we have $N$ particles which have spin
projections on certain axis equal to either $-\frac{1}{2}$ or $+\frac{1}{2}$.
The microscopic state of such system is fully described by a set of spin
projections $\{s_{1},s_{2},\ldots,s_{N}\}$. This system may be treated, in the
statistical mechanics sense, using the microcanonical ensemble in which each of
the microscopic states is assigned the same probability. Here we obtain the same
stationary probability density function (PDF) from the temporal dynamics
perspective.

We are interested in an observable macroscopic quantity, the total spin of the
system
\begin{equation}
M=\sum_{i=1}^{N}s_{i}\,.\label{eq:total-spin}
\end{equation}
Alternatively we can express the total spin as
\begin{equation}
M=\frac{1}{2}(N_{+}-N_{-})\,,\label{eq:m-np-nm}
\end{equation}
where $N_{+}$ is the number of spins with projection $+\frac{1}{2}$ and $N_{-}$
is the number of spins with projection $-\frac{1}{2}$. A sum of $N_{+}$ and
$N_{-}$, by definition, should give the total number of spins $N$,
\begin{equation}
N_{+}+N_{-}=N\,.\label{eq:n-np-nm}
\end{equation}
From Eqs.~(\ref{eq:m-np-nm}) and (\ref{eq:n-np-nm}) we obtain relation between
total spin and number of particles having certain spin projections:
\begin{equation}
N_{+}=\frac{1}{2}N+M\,,\qquad N_{-}=\frac{1}{2}N-M\,.\label{eq:np-nm}
\end{equation}

To introduce temporal dynamics we assume that during the short time interval
$\Delta t$ each spin can flip with the probability $\frac{1}{2}\gamma\Delta t$.
If $\Delta t$ is short enough we can assume that only one spin flip takes place
during the time interval $\Delta t$. Since each spin can flip independently, the
transition probabilities per unit time are
\begin{eqnarray}
p(M\rightarrow M+1) & \equiv & p^{+}(M)=\frac{1}{2}\gamma N_{-}\,,
\label{eq:prob-p}\\
p(M\rightarrow M-1) & \equiv & p^{-}(M)=\frac{1}{2}\gamma N_{+}\,.
\label{eq:prob-m}
\end{eqnarray}
The transition probabilities $p^{+}(M)$ and $p^{-}(M)$ define a one-step
stochastic process and imply the following Master equation for the probability
$P_{M}(M,t)$ to find the value $M$ of the total spin at time $t$
\cite{Kampen2011}:
\begin{equation}
\frac{\partial}{\partial t}P_{M}(M,t)=p^{+}(M-1)P_{M}(M-1,t)
+p^{-}(M+1)P_{M}(M+1,t)-(p^{+}(M)+p^{-}(M))P_{M}(M,t)\,.
\label{eq:master}
\end{equation}
For large enough $N$ we can represent the dynamics by a continuous variable
$x=2M/N$. Using the standard methods from Ref.~\cite{Kampen2011} one can derive
the Fokker-Planck equation from the Master equation (\ref{eq:master}) assuming
that $N$ is large and neglecting the terms of the order of $1/N^{2}$. The
resulting Fokker-Planck equation is
\begin{equation}
\frac{\partial}{\partial t}P_{x}(x,t)=-\frac{\partial}{\partial x}
[\pi^{+}(x)-\pi^{-}(x)]P_{x}(x,t)
+\frac{\partial^{2}}{\partial x^{2}}\frac{1}{N}[\pi^{+}(x)+\pi^{-}(x)]P_{x}(x,t)\,,
\label{eq:FP-general}
\end{equation}
where $P_{x}(x,t)=(N/2)P_{M}(Nx/2,t)$ is the PDF of the stochastic variable $x$
and
\begin{equation}
\pi^{\pm}(x)\equiv\frac{2}{N}p^{\pm}\left(\frac{Nx}{2}\right)\,.
\end{equation}
The SDE corresponding to the Fokker-Planck equation (\ref{eq:FP-general}) is
\cite{Gardiner2009}
\begin{equation}
\frac{dx}{dt}=[\pi^{+}(x)-\pi^{-}(x)]+\sqrt{\frac{2}{N}[\pi^{+}(x)+\pi^{-}(x)]}\xi(t)\,.
\label{eq:SDE-general}
\end{equation}
Here $\xi(t)$ is a white noise with autocorrelation
$\langle\xi(t)\xi(t')\rangle=\delta(t-t')$. When the diffusion coefficient
depends on $x$, SDE (\ref{eq:SDE-general}) should be understood in It\^o
convention.

For the transition probabilities (\ref{eq:prob-p}) and (\ref{eq:prob-m}), the
Fokker-Planck equation (\ref{eq:FP-general}) becomes
\begin{equation}
\frac{\partial}{\partial t}P_{x}(x,t)=\gamma\frac{\partial}{\partial x}xP_{x}(x,t)
+\frac{\gamma}{N}\frac{\partial^{2}}{\partial x^{2}}P_{x}(x,t)
\label{eq:FP-1}
\end{equation}
and Eq.~(\ref{eq:SDE-general}) takes the form
\begin{equation}
\frac{dx}{dt}=-\gamma x+\sqrt{\frac{2\gamma}{N}}\xi(t)\,.\label{eq:sde-1}
\end{equation}
As one can see from Eq.~(\ref{eq:sde-1}), the fluctuations of the macroscopic
quantity $M$ decrease with the increasing number of spins as $1/\sqrt{N}$ . From
Eq.~(\ref{eq:sde-1}) follows that the average value of $x$ obeys the ordinary
differential equation
\begin{equation}
\frac{d}{dt}\langle x\rangle=-\gamma\langle x\rangle\,.
\end{equation}
Thus the deviations from the equilibrium average $\langle x\rangle_{0}=0$ decay
exponentially: $\langle x(t)\rangle=\langle x(0)\rangle e^{-\gamma t}$. The
steady-state PDF of the stochastic variable $x$ obtained from the Fokker-Planck
equation (\ref{eq:FP-1}) is Gaussian,
\begin{equation}
P_{0}(x)=\sqrt{\frac{N}{2\pi}}\exp\left(-\frac{N}{2}x^{2}\right)\,.
\label{eq:uncor-gauss}
\end{equation}
Gaussian distribution, coinciding with Eq.~(\ref{eq:uncor-gauss}), is also
obtained assuming equal probabilities for each microscopic configuration.

\section{Stochastic dynamics in the model with correlated binary variables}

\label{sec:correlated} In this Section we investigate the statistical model
consisting of $N$ correlated binary random variables, similar to the model of
Ref.~\cite{Ruseckas2015}. As in the previous Section one can consider particles
having spins $\frac{1}{2}$. In order to avoid special treatment of the ends we
will consider the spins situated on a ring instead of one-dimensional chain as
in Ref~\cite{Ruseckas2015}. We assume that the spins are correlated, meaning
that the two adjacent spins are aligned, except for the $d$ cases when the two
adjacent spins are antialigned. Note that for a ring of spins only even $d$ is
possible. Thus the ring consists of $d$ domains with aligned spins and has $d$
domain boundaries where the spin flips occur. Similarly as in
Ref~\cite{Ruseckas2015} one can show that the number of allowed microscopic
configurations in the model grows with the number $N$ of spins as $N^{d}$ and
the equilibrium distribution of the total spin $M$ in the limit
$N\rightarrow\infty$ is a $q$-Gaussian with $q_{\mathrm{dist}}=1-2/(d-2)$.

We introduce stochastic temporal dynamics into the model similarly as we have
done in Section~\ref{sec:uncorrelated}: we assume that during the short time
interval $\Delta t$ spins can flip with the probability $\frac{1}{2}\gamma\Delta
t$. However, in order to conserve $d$, only the spins next to the domain
boundaries are allowed to flip. Furthermore, in order to conserve $d$, spins
belonging to domains containing single spin should not be allowed to flip. If
such spin would be allowed to flip, the domains could disappear and $d$ would
not be conserved. Alternatively, these spin-flip rules may be seen as
introducing a random walk of the domain boundaries. Namely, during the short
time interval $\Delta t$ each domain boundary can move to the left or to the
right with equal probabilities $\frac{1}{2}\gamma\Delta t$ unless the move
results in two boundaries in the same position. As in
Section~\ref{sec:uncorrelated} we will obtain the Fokker-Plank equation
describing how the distribution of the total spin changes in time.

If the time interval$\Delta t$ is short enough, we may assume that only one spin
flip takes place at the time. Since each boundary may flip independently, in the
case with no one-spin domains the transition probabilities per unit time are
given by 
\begin{equation}
p(M\rightarrow M+1)=p(M\rightarrow M-1)=\frac{1}{2}\gamma d\,.
\end{equation}
Evidently the presence of one-spin domains would make spin flip less probable.
Given only the number of spins $N$ and the total spin $M$ the number of one-spin
domains is not exactly known, therefore the actual transition probabilities
corresponding to the same total spin $M$ will change over time. For the
analytical description of the model we will use averaged in time transition
probabilities corresponding to given $M$. If $K_{+}(M)$ is the average number of
one-spin domains with spin projection $+\frac{1}{2}$ and $K_{-}(M)$ is the
average number of one-spin domains with spin projection $-\frac{1}{2}$, then the
average transition probabilities per unit time are
\begin{eqnarray}
\langle p(M\rightarrow M+1)\rangle & = & \frac{1}{2}\gamma(d-2K_{-})\,.
\label{eq:prob-p-2}\\
\langle p(M\rightarrow M-1)\rangle & = & \frac{1}{2}\gamma(d-2K_{+})\,.
\label{eq:prob-m-2}
\end{eqnarray}

We will assume that during the temporal evolution the probabilities of different
microscopic configurations are almost equal. Then the average number of one-spin
domains can be expressed as
\begin{equation}
K_{\pm}=\sum_{i=0}^{\frac{d}{2}}i\frac{W_{i}\left(N_{\pm},\frac{d}{2}\right)}{W_{\mathrm{div}}
\left(N_{\pm},\frac{d}{2}\right)}\,,
\label{eq:average-k}
\end{equation}
where $W_{\mathrm{div}}(N_{\pm},d/2)$ is the number of possible divisions of
$N_{\pm}$ spins into $d/2$ domains and $W_{i}(N_{\pm},d/2)$ is the number of
possible divisions of $N_{\pm}$ spins into $d/2$ domains such that there are $i$
one-spin domains. The number $N_{+}$ of spins with projection $+\frac{1}{2}$ and
the number $N_{-}$ of spins with projection $-\frac{1}{2}$ may be obtained from
the known $N$ and $M$ using Eq.~(\ref{eq:np-nm}). The spins with projection
$+\frac{1}{2}$ as well as the spins with projection $-\frac{1}{2}$ are divided
into $d/2$ domains. Thus the number of possible divisions of spins with the same
projection into domains is
\begin{equation}
W_{\mathrm{div}}\left(N_{\pm},\frac{d}{2}\right)=\binom{N_{\pm}-1}{\frac{d}{2}-1}\,.
\end{equation}
Here $\binom{m}{n}$ is the binomial coefficient. The number of possible
configurations with $i$ one-spin domains can be expressed as 
\begin{equation}
W_{i}\left(N_{\pm},\frac{d}{2}\right)=\binom{\frac{d}{2}}{i}W_{0}
\left(N_{\pm}-i,\frac{d}{2}-i\right)\,.
\label{eq:i-one-dom}
\end{equation}
Here the binomial coefficient $\binom{d/2}{i}$ gives the number of ways to
choose $i$ one-spin domains from $d/2$ domains, while $W_{0}(N_{\pm}-i,d/2-i)$
is the number of ways to distribute the remaining $N_{\pm}-i$ spins into the
remaining $d/2-i$ domains containing more than one spin. We obtain the
expression for the number $W_{0}(n,k)$ of possible divisions of $n$ spins into
$k$ domains with no one-spin domains as follows. First we put aside $k$ spins.
Next we distribute the remaining $n-k$ spins into $k$ domains with at least one
spin each. Afterwards we add one spin to each of the $k$ domains using the spins
put aside in the first step. Evidently there is only one way to complete the
last step, thus the number $W_{0}(n,k)$ is determined by the number of ways to
divide $n-k$ spins into $k$ domains: 
\begin{equation}
W_{0}(n,k)=\binom{n-1-k}{k-1}\,.\label{eq:0-one-dom}
\end{equation}
Using Eqs.~(\ref{eq:average-k})--(\ref{eq:0-one-dom}) we obtain the average
number of one-spin domains
\begin{equation}
K_{\pm}=\frac{d}{2}\frac{\frac{d}{2}-1}{N_{\pm}-1}\,.\label{eq:avg-one-spin}
\end{equation}
Inserting Eq.~(\ref{eq:avg-one-spin}) into Eqs.~(\ref{eq:prob-p-2}) and
(\ref{eq:prob-m-2}) we get the average transition probabilities per unit time
\begin{eqnarray}
\langle p(M\rightarrow M+1)\rangle & = & \frac{\gamma d}{2}\frac{N_{-}-\frac{d}{2}}{N_{-}-1}\,,
\label{eq:prob-p-3}\\
\langle p(M\rightarrow M-1)\rangle & = & \frac{\gamma d}{2}\frac{N_{+}-\frac{d}{2}}{N_{+}-1}\,.
\label{eq:prob-m-3}
\end{eqnarray}
Using Eqs.~(\ref{eq:prob-p-3}) and (\ref{eq:prob-m-3}) in
Eq.~(\ref{eq:FP-general}) and keeping the terms of the order of $1/N^{2}$ we
obtain the Fokker-Planck equation 
\begin{equation}
\frac{\partial}{\partial t}P_{x}(x,t)=\frac{4\gamma d}{N^{2}}\left(\frac{d}{2}-1\right)
\frac{\partial}{\partial x}\frac{x}{1-x^{2}}P_{x}(x,t)+\frac{2\gamma d}{N^{2}}
\frac{\partial^{2}}{\partial x^{2}}P_{x}(x,t)\,.
\label{eq:FP-2}
\end{equation}
The corresponding SDE, according to Eq.~(\ref{eq:SDE-general}), is
\begin{equation}
\frac{dx}{dt}=-\frac{4\gamma d}{N^{2}}\left(\frac{d}{2}-1\right)\frac{x}{1-x^{2}}
+\frac{2}{N}\sqrt{\gamma d}\xi(t)\,.
\label{eq:sde-2}
\end{equation}
The dependence on the number of spins $N$ in SDE (\ref{eq:sde-2}) can be removed
by introducing the scaled time $t_{\mathrm{s}}=\frac{4\gamma d}{N^{2}}t$:
\begin{equation}
\frac{dx}{dt_{\mathrm{s}}}=-\left(\frac{d}{2}-1\right)\frac{x}{1-x^{2}}+\xi(t_{\mathrm{s}})\,.
\label{eq:sde-3}
\end{equation}
From Eq.~(\ref{eq:sde-2}) follows that the evolution of the total spin slows
down with increasing number of spins $N$ as $1/N^{2}$. This is in contrast to
the case of uncorrelated spins, Eq.~(\ref{eq:sde-1}), where the relaxation
towards equilibrium does not depend on $N$. For the model of correlated spins
the fluctuations of the total spin remain macroscopic even in the limit
$N\rightarrow\infty$, the size of fluctuations of $x$ does not depend on the
number of spins $N$. This is different than in the model of uncorrelated spins
where the size of fluctuations decreases as $1/\sqrt{N}$. Steady-state PDF of
the stochastic variable $x$ obtained from the Fokker-Planck equation
(\ref{eq:FP-2}) is
\begin{equation}
P_{0}(x)\propto(1-x^{2})^{\frac{d}{2}-1}\,.
\end{equation}
This steady-state PDF has a $q$-Gaussian form with $q<1$ and coincides with the
distribution of the total spin obtained assuming equal probabilities for each
microscopic configuration. The Fokker-Planck equation (\ref{eq:FP-2}) is a
particular case of known Fokker-Planck equations giving $q$-Gaussian
distributions and satisfies the condition given by Eq.~(11) of
Ref.~\cite{Borland1998-2}.

\subsection{Testing the Markovian approximation of the dynamical model with
correlated binary variables}

\begin{figure}
\includegraphics[width=0.45\textwidth]{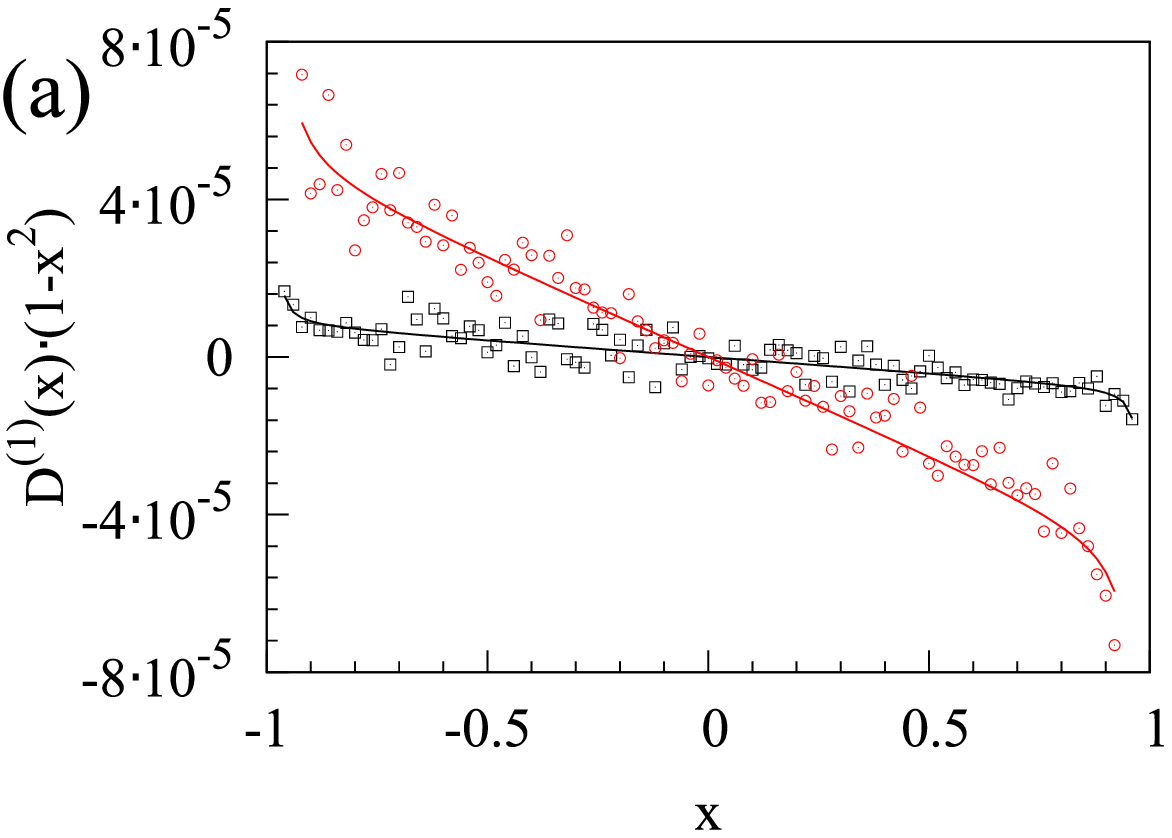}\includegraphics[width=0.45\textwidth]{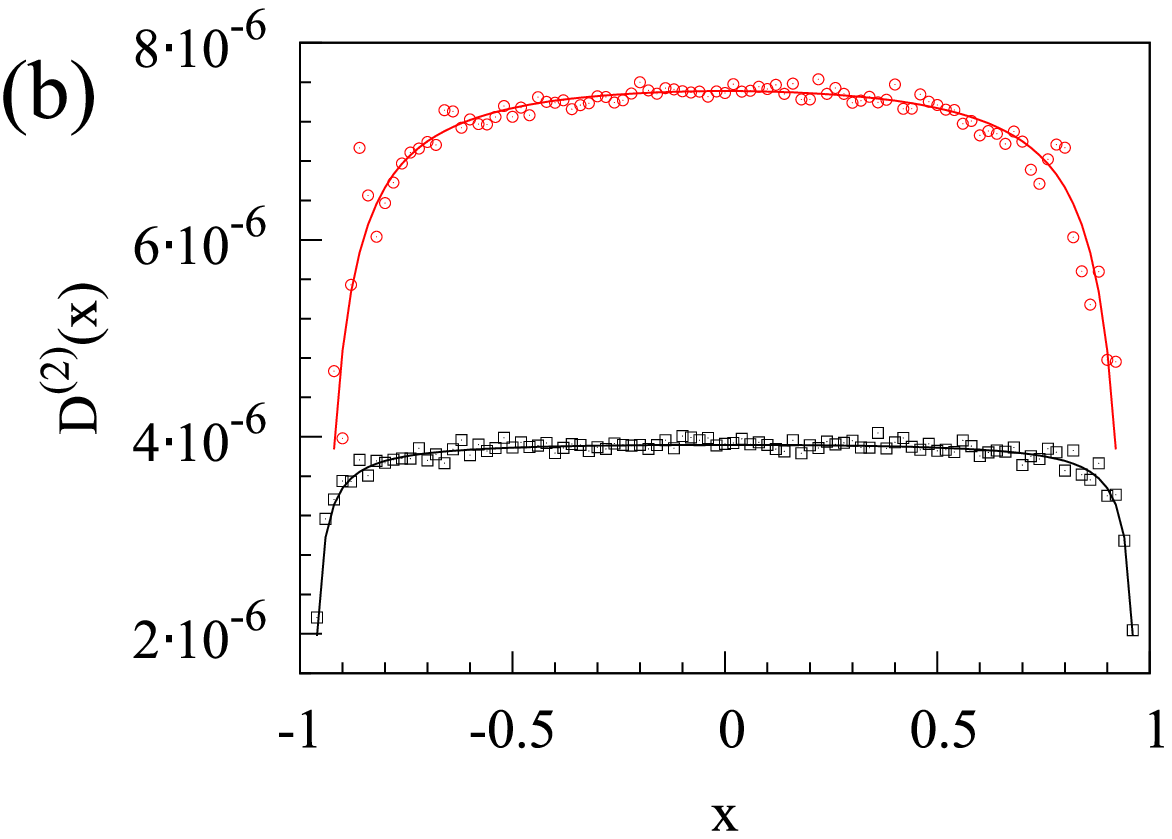}
\caption{Kramers-Moyal coefficients (a) $D^{(1)}(x)$ and (b) $D^{(2)}(x)$
(symbols) for two different numbers of spin flips estimated using numerical time
series. In order to get a curve closer to a straight line the coefficient
$D^{(1)}(x)$ is multiplied by $1-x^{2}$. Solid lines show corresponding analytic
expressions (\ref{eq:d1}) and (\ref{eq:d2}). The number of spins is $N=100$, the
number of spin flips is $d=4$ (black squares) and $d=8$ (red circles). Other
parameters are $\gamma=0.005$, $\Delta t=1$.}
\label{fig:Kramers-Moyal}
\end{figure}

\begin{figure}
\includegraphics[width=0.45\textwidth]{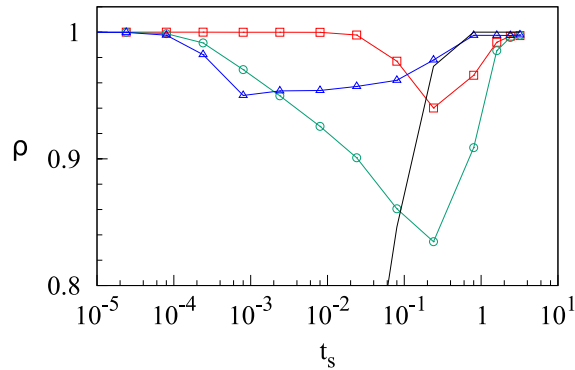}
\caption{Dependence of the Pearson product-moment correlation coefficient $\rho$
on the scaled time $t_{\mathrm{s}}$. The coefficient $\rho$ is calculated
between the time-dependent PDF obtained from the Markovian approximation and the
numerically obtained time-dependent PDFs when the initial configuration contains
no one-spin domains (red squares), one one-spin domain (green circles), two
one-spin domains (blue triangles). The solid curves passing through the symbols
are for the convenience of the eyes only. Black solid curve without symbols
shows the Pearson product-moment correlation coefficient between the
time-dependent PDF obtained from the Markovian approximation and the
steady-state PDF. Model parameters are $N=100$, $d=4$, $\gamma=0.005$, $\Delta
t=1$.}
\label{fig:pearson}
\end{figure}

To check the assumptions made in deriving the Fokker-Planck equation
(\ref{eq:FP-2}) we numerically simulate the temporal evolution of the ring of
spins and obtain the time series describing the time dependence of the total
spin $M$. Using those time series we estimate Kramers-Moyal coefficients and
compare them to the analytical predictions. In addition we compare
time-dependent PDFs of the total spin to the PDFs obtained using Markovian
approximation with the transition probabilities per unit time
(\ref{eq:prob-p-3}), (\ref{eq:prob-m-3}).

In numerical simulation we have chosen a random initial configuration with
$N=100$ spins and $d=4$ or $d=8$ spin flips, a small time step $\Delta t$ such
that $\gamma\Delta t\ll1$ and generated time series with $3\cdot10^{7}$ points.
Using the numerical time series we estimate Kramers-Moyal coefficients
\cite{Risken1996} 
\begin{equation}
D^{(n)}(X)=\frac{1}{n!\Delta t}\left.\langle[x(t+\Delta t)-x(t)]^{n}\rangle\right|_{x(t)=X}\,,
\end{equation}
where $x=2M/N$. The numerically obtained Kramers-Moyal coefficients $D^{(1)}(x)$
and $D^{(2)}(x)$ are shown as symbols in Fig.~\ref{fig:Kramers-Moyal}. We have
found the coefficients $D^{(3)}$ and $D^{(4)}$ to be $4$ to $5$ degrees of
magnitude smaller than $D^{(1)}$ and $D^{(2)}$. Thus, according to the Pawula
theorem the dynamics of the complete model may be approximated by an
Fokker-Planck equation. The continuous curves in Fig.~\ref{fig:Kramers-Moyal}
show the analytical expressions for the Kramers-Moyal coefficients
\begin{eqnarray}
D^{(1)}(x) & = & -\frac{4\gamma d}{N^{2}}
\frac{\left(\frac{d}{2}-1\right)x}{\left(1-\frac{2}{N}\right)^{2}-x^{2}}\,,
\label{eq:d1}\\
D^{(2)}(x) & = & \frac{2\gamma d}{N^{2}}\frac{\left(1-\frac{d}{N}\right)
\left(1-\frac{2}{N}\right)-x^{2}}{\left(1-\frac{2}{N}\right)^{2}-x^{2}}\,,
\label{eq:d2}
\end{eqnarray}
that take into account the effects of the finite number of the spins. To derive
Eqs.~(\ref{eq:d1}), (\ref{eq:d2}) we have inserted Eqs.~(\ref{eq:prob-p-3}) and
(\ref{eq:prob-m-3}) into Eq.~(\ref{eq:FP-general}) without making any further
approximations. We see in Fig.~\ref{fig:Kramers-Moyal} a good agreement of the
numerically estimated Kramers-Moyal coefficients with the analytic predictions.
In the limit of $N\rightarrow\infty$ the curves in Fig.~\ref{fig:Kramers-Moyal}
should be straight lines. However, as Eqs.~(\ref{eq:d1}), (\ref{eq:d2}) and the
numerical simulations show, there are deviations from the straight lines for the
most extreme values of the total spin. The range of $x$ values where the
deviations are significant decreases with increasing $N$.

Total spin of the system does not determine the transition probabilities
completely, because these probabilities depend also on the number of one-spin
domains. To check the impact of this missing information on the temporal
evolution of the system we have randomly generated several initial spin
configurations, each with $M=0$, and observed how well the numerically obtained
time-dependent PDFs (calculated using $10^{5}$ realizations that start from same
initial configuration and contain $5\cdot10^{5}$ points each) correspond to the
time-dependent PDF obtained from the Markovian approximation. The dependence of
the Pearson product-moment correlation coefficient $\rho$ on the scaled time
$t_{\mathrm{s}}$ is shown in Fig.\ \ref{fig:pearson}. One can see that the
differences in initial spin configuration play a significant role for
intermediate times $t_{\mathrm{s}}\lesssim1$ as the Pearson product-moment
correlation coefficient becomes smaller. The initial configuration becomes less
important as the time-dependent PDF approaches the steady-state PDF.

\section{Long-range temporal dependence in the model}

\label{sec:long-range}Presence of macroscopic fluctuations in the model
presented in previous Section can enable recovery of $1/f$ noise. To demonstrate
this let us consider a quantity equal to the ratio of the the number of spins
with projection $-\frac{1}{2}$ to the number of spins with projection
$+\frac{1}{2}$:
\begin{equation}
y=\frac{N_{-}}{N_{+}}=\frac{1-x}{1+x}\,.
\end{equation}
The range of possible $y$ values is $[d/(2N-d),(2N-d)/d]$. Using It\^o's lemma
one may obtain the following SDE for the stochastic variable $y$
\begin{equation}
\frac{dy}{dt_{\mathrm{s}}}=\frac{1}{4}\left(2-\frac{\lambda_{1}}{2}
+\frac{\lambda_{2}}{2y}\right)(1+y)^{3}+\frac{1}{2}(1+y)^{2}\xi(t_{\mathrm{s}})\,,
\label{eq:sde-y}
\end{equation}
where
\begin{equation}
\lambda_{1}=\frac{d}{2}+1\,,\qquad\lambda_{2}=\frac{d}{2}-1\,.
\end{equation}
The steady-state PDF of the stochastic variable $y$, obtained from the
Fokker-Planck equation corresponding to the SDE (\ref{eq:sde-y}), is
\begin{equation}
P_{0}(y)=\frac{\Gamma(\lambda_{1}+\lambda_{2})}{\Gamma(\lambda_{1}-1)\Gamma(\lambda_{2}+1)}
\frac{y^{\lambda_{2}}}{(1+y)^{\lambda_{1}+\lambda_{2}}}=
\frac{\Gamma(\lambda_{1}+\lambda_{2})}{\Gamma(\lambda_{1}-1)\Gamma(\lambda_{2}+1)}
\exp_{q_{2}}\left(-\frac{\lambda_{2}}{y}\right)\exp_{q_{1}}(-\lambda_{1}y)\,,
\label{eq:PDF-y}
\end{equation}
where
\begin{equation}
q_{1}=1+1/\lambda_{1}\,,\qquad q_{2}=1+1/\lambda_{2}\,.
\end{equation}
Thus the steady-state PDF of $y$ is a $q$-exponential with $q$-exponential
cut-off at small values of $y$. The SDE (\ref{eq:sde-y}) for large $y\gg1$
coincides with the nonlinear SDE proposed in
Refs.~\cite{Kaulakys2004,Kaulakys2006} (with the power-law exponent in the drift
term $\eta=2$). This similarity implies that SDE (\ref{eq:sde-y}) should
generate time series exhibiting power spectral density (PSD) of
$S(f)\sim1/f^{\beta}$ form, where
\begin{equation}
\beta=1+\frac{\lambda_{1}-3}{2(\eta-1)}=\frac{d}{4}\,.
\end{equation}
This expression is valid for $0<\beta\leqslant2$. Thus we may expect that the
PSD of the ratio $N_{-}/N_{+}$ should be $1/f$ in a wide range of frequencies
when $d=4$. The range of frequencies is limited by the finite number of spins
(the ratio $N_{-}/N_{+}$ has a finite maximum possible value
$y_{\mathrm{max}}\approx2N/d$) as well as by the steady-state PDF of $y$
exhibiting power-law behavior with the power-law exponent $\lambda_{1}$ only for
$y\gg1$. According to Ref.~\cite{Ruseckas2014} the frequencies in the power-law
part of the PSD satisfy $\sigma^{2}y_{\mathrm{min}}^{2(\eta-1)}\ll2\pi
f_{\mathrm{s}}$, where $\sigma$ is the coefficient in the noise term. For SDE
(\ref{eq:sde-y}) $y_{\mathrm{min}}=1$ and $\sigma=\frac{1}{2}$. Going back from
the scaled time $t_{\mathrm{s}}$ to the physical time $t$ we get that the PSD
has power-law behavior for frequencies
\begin{equation}
\frac{\gamma d}{N^{2}}\ll2\pi f\,.\label{eq:f-min}
\end{equation}
We see that the lowest limiting frequency decreases with increase of the number
of spins as $1/N^{2}$. The width of the frequency region with the power-law
behavior of the PSD grows with increasing particle number $N$.

\begin{figure}
\includegraphics[width=0.45\textwidth]{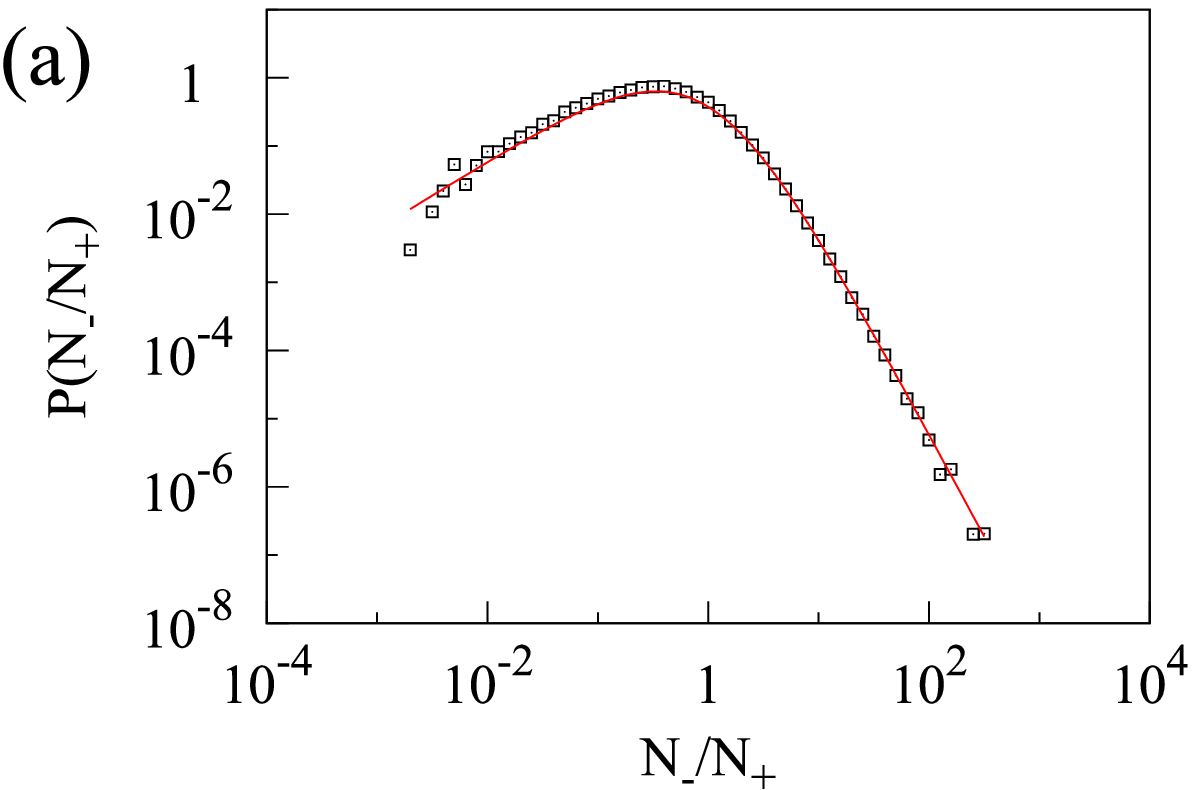}\includegraphics[width=0.45\textwidth]{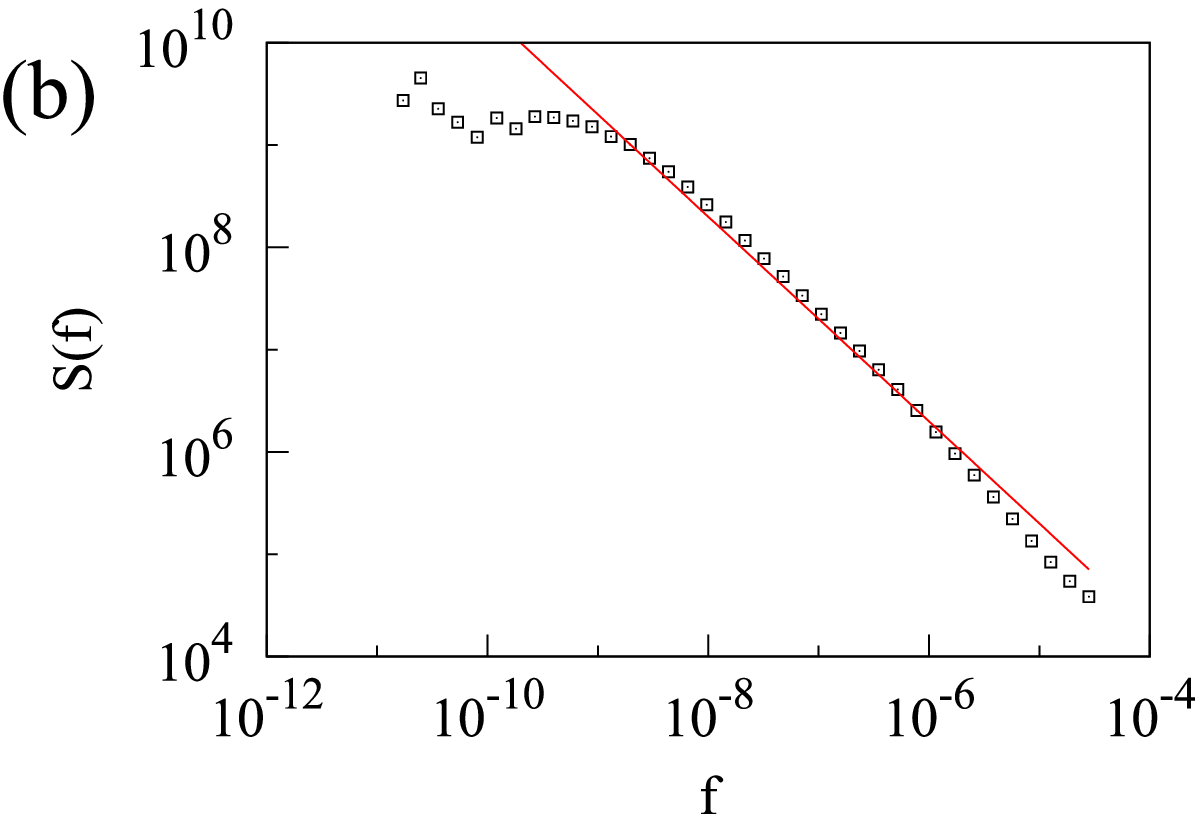}
\caption{(a) The steady-state PDF of the ratio $N_{-}/N_{+}$. Solid (red) line
corresponds to Eq.~(\ref{eq:PDF-y}). (b) The PSD of the ratio $N_{-}/N_{+}$.
Solid (red) line shows the slope $f^{-1}$. Black squares show numerical results
obtained using simulation of the stochastic evolution of the ring of spins.
Model parameters are $\gamma=0.005$, $N=10^{3}$, $d=4$.}
\label{fig:spectum}
\end{figure}

The power-spectral density of the ratio $y=N_{-}/N_{+}$ obtained from numerical
simulation of the stochastic evolution of the ring of spins when $d=4$ is shown
in Fig.~\ref{fig:spectum}. We see a good agreement of the numerical results with
the analytical predictions. The numerical simulation confirms the presence of
$1/f$ region in the spectrum of the ratio $N_{-}/N_{+}$. The $1/f$ interval in
the PSD in Fig.~\ref{fig:spectum}(b) is approximately between
$f_{\mathrm{min}}\approx10^{-9}$ and $f_{\mathrm{max}}\approx10^{-6}$, the lowest
frequency $f_{\mathrm{min}}$ is in agreement with the estimation
(\ref{eq:f-min}). The width of this frequency interval can be increased by
increasing $N$.

\section{Interacting subsystems}

\label{sec:subsyst}The temporal dynamics in the model of correlated spins,
proposed in Section~\ref{sec:correlated}, allows us to consider a contact
between two such systems. Since the dynamics results in a random walk of the
spin flip positions, it is natural to assume that the contact between two
systems allows for spin flips to move from one system to the other. If the first
system has $d_{1}$ spin flips, the second system has $d_{2}$ spin flips, then
the total number of spin flips in the composite system is $d=d_{1}+d_{2}$.
Taking into account the relation $q_{\mathrm{stat}}=1-1/d$ \cite{Ruseckas2015},
we obtain that the index $q_{\mathrm{stat}}$ of the composite system obeys the
equation
\begin{equation}
\frac{1}{1-q_{\mathrm{stat}}}=\frac{1}{1-q_{\mathrm{stat},1}}
+\frac{1}{1-q_{\mathrm{stat},2}}\,.
\end{equation}
Note, that for such composite system the number of allowed microscopic
configurations $W$ is not equal to the product $W_{1}W_{2}$, where $W_{i}$ is
the number of allowed microscopic configurations of the isolated $i$-th system.
Since the number of allowed microscopic configurations grows as $W\sim N^{d}$
\cite{Ruseckas2015} and for composite system $N=N_{1}+N_{2}$ with
$d=d_{1}+d_{2}$, we have
\begin{equation}
W\sim(W_{1}^{1-q_{\mathrm{stat},1}}+W_{2}^{1-q_{\mathrm{stat},2}})^{\frac{1}{1-q_{\mathrm{stat}}}}\,.
\end{equation}
This equation generalizes Eq.~(3.113) of Ref.~\cite{Tsallis2009-1} for
subsystems with different values of $q$.

The subsystem of larger composite system has fluctuating number of spin flips.
Let us calculate the probability distribution of spin flips in the subsystems
when this distribution becomes stationary. In this situation we can assume that
each microscopic configuration of the composite system has the same probability.
The number of ways to partition a ring of $N$ spins into $d$ domains is equal to
the number of ways $\binom{N}{d}$ to place $d$ domain boundaries into $N$
possible positions. This number should be multiplied by $2$ because there are
$2$ ways to assign the signs of spin projections to the domains. Thus the number
of allowed microscopic configurations of the ring of spins is
\begin{equation}
W=2\binom{N}{d}\label{eq:w-total}
\end{equation}
Let us consider a subsystem as a spin chain having $N_{1}$ spins being a part of
a larger ring with $N$ spins. The number of microscopic configurations where the
subsystem has $d_{1}$domain boundaries is proportional to the number of ways to
place $d_{1}$ domain boundaries into $N_{1}-1$ positions in the subsystem
multiplied by the number of ways to place remaining $d-d_{1}$ domain boundaries
into remaining $N-N_{1}+1$ positions:
\begin{equation}
W(d_{1})=2\binom{N_{1}-1}{d_{1}}\binom{N-N_{1}+1}{d-d_{1}}\label{eq:w-d1}
\end{equation}
The probability to have $d_{1}$ boundaries in the subsystem is equal to the
ratio
\begin{equation}
P(d_{1})=\frac{W(d_{1})}{W}\,.\label{eq:prob-d1}
\end{equation}
Note that the probabilities are normalized, $\sum_{d_{1}=0}^{d}P(d_{1})=1$.
Using Eqs.~(\ref{eq:w-total})--(\ref{eq:prob-d1}) for large $N_{1}\gg d_{1}$ and
$N\gg d$ we obtain the distribution of the spin flips in the subsystem 
\begin{equation}
P(d_{1})\approx\frac{d!}{(d-d_{1})!d_{1}!}\left(\frac{N_{1}}{N-N_{1}}\right)^{d_{1}}
\left(\frac{N-N_{1}}{N}\right)^{d}\,.
\end{equation}
The average number of spin flips in the subsystem calculated using the
probabilities (\ref{eq:prob-d1}) is 
\begin{equation}
\langle d_{1}\rangle=d\frac{N_{1}}{N}\,.
\end{equation}
Thus for the two subsystems in contact we have the equality
\begin{equation}
\frac{\langle d_{1}\rangle}{N_{1}}=\frac{\langle d_{2}\rangle}{N_{2}}
\end{equation}
when distributions of spin flips in the subsystems become stationary. We can
interpret the quantity $\Theta=\langle d_{1}\rangle/N_{1}$ as an effective
temperature. An effective temperature associated with the non-extensive
statistical mechanics and proportional to the density of vortices in type II
superconductors has been introduced in \cite{Nobre2012,Nobre2015}.

\section{Conclusions}

\label{sec:conclusions} We have extended the statistical model proposed in
Ref.~\cite{Ruseckas2015} that exhibits both extensive behavior of generalized
entropy for $q<1$ and a $q$-Gaussian distribution. Assuming that the stochastic
temporal dynamics is due to random walk of domain boundaries we have derived the
Fokker-Planck equation (\ref{eq:FP-2}) and corresponding SDE (\ref{eq:sde-2})
describing the evolution of the total spin in time. Although distributions from
non-extensive statistical mechanics are more often obtained using nonlinear
Fokker-Planck equations \cite{Tsallis2009-1}, our model is well described by a
linear Fokker-Planck equation. Eq.~(\ref{eq:FP-2}) is a particular case of a
known class of Fokker-Planck equations, which have $q$-Gaussian stationary
distributions distributions \cite{Borland1998-2}. The proposed temporal dynamics
in the model provides insight on how Fokker-Planck equations with stationary
distributions of non-extensive statistical mechanics can arise from microscopic
description of the system.

In contrast to the model with uncorrelated spins, presented in Section
\ref{sec:uncorrelated}, the dynamics in the proposed model of correlated spins
slow down with the increasing number of spins as $1/N^{2}$. In addition, the
fluctuations of the total spin remain macroscopic even in the limit
$N\rightarrow\infty$, whereas in the model with uncorrelated spins the
fluctuations decrease as $1/\sqrt{N}$. These macroscopic fluctuations is one of
the reasons for the spectrum of fluctuations of the ratio $N_{-}/N_{+}$
exhibiting $1/f$ behavior in a wide range of frequencies. This range of
frequencies increases with increasing number of spins $N$.

The model presented in this Letter works for $q<1$. Thus the question remains
whether it is possible to modify the model to obtain $q$-Gaussian distribution
with $q>1$.


\begin{thebibliography}{36}%
\makeatletter
\providecommand \@ifxundefined [1]{%
 \@ifx{#1\undefined}
}%
\providecommand \@ifnum [1]{%
 \ifnum #1\expandafter \@firstoftwo
 \else \expandafter \@secondoftwo
 \fi
}%
\providecommand \@ifx [1]{%
 \ifx #1\expandafter \@firstoftwo
 \else \expandafter \@secondoftwo
 \fi
}%
\providecommand \natexlab [1]{#1}%
\providecommand \enquote  [1]{``#1''}%
\providecommand \bibnamefont  [1]{#1}%
\providecommand \bibfnamefont [1]{#1}%
\providecommand \citenamefont [1]{#1}%
\providecommand \href@noop [0]{\@secondoftwo}%
\providecommand \href [0]{\begingroup \@sanitize@url \@href}%
\providecommand \@href[1]{\@@startlink{#1}\@@href}%
\providecommand \@@href[1]{\endgroup#1\@@endlink}%
\providecommand \@sanitize@url [0]{\catcode `\\12\catcode `\$12\catcode
  `\&12\catcode `\#12\catcode `\^12\catcode `\_12\catcode `\%12\relax}%
\providecommand \@@startlink[1]{}%
\providecommand \@@endlink[0]{}%
\providecommand \url  [0]{\begingroup\@sanitize@url \@url }%
\providecommand \@url [1]{\endgroup\@href {#1}{\urlprefix }}%
\providecommand \urlprefix  [0]{URL }%
\providecommand \Eprint [0]{\href }%
\providecommand \doibase [0]{http://dx.doi.org/}%
\providecommand \selectlanguage [0]{\@gobble}%
\providecommand \bibinfo  [0]{\@secondoftwo}%
\providecommand \bibfield  [0]{\@secondoftwo}%
\providecommand \translation [1]{[#1]}%
\providecommand \BibitemOpen [0]{}%
\providecommand \bibitemStop [0]{}%
\providecommand \bibitemNoStop [0]{.\EOS\space}%
\providecommand \EOS [0]{\spacefactor3000\relax}%
\providecommand \BibitemShut  [1]{\csname bibitem#1\endcsname}%
\let\auto@bib@innerbib\@empty
\bibitem [{\citenamefont {Tsallis}(2009{\natexlab{a}})}]{Tsallis2009-1}%
  \BibitemOpen
  \bibfield  {author} {\bibinfo {author} {\bibfnamefont {C.}~\bibnamefont
  {Tsallis}},\ }\href@noop {} {\emph {\bibinfo {title} {Introduction to
  Nonextensive Statistical Mechanics---Approaching a Complex World}}}\
  (\bibinfo  {publisher} {Springer},\ \bibinfo {address} {New York},\ \bibinfo
  {year} {2009})\BibitemShut {NoStop}%
\bibitem [{\citenamefont {Gell-Mann}\ and\ \citenamefont
  {Tsallis}(2004)}]{Gell-Mann2004}%
  \BibitemOpen
  \bibfield  {author} {\bibinfo {author} {\bibfnamefont {C.~M.}\ \bibnamefont
  {Gell-Mann}}\ and\ \bibinfo {author} {\bibfnamefont {C.}~\bibnamefont
  {Tsallis}},\ }\href@noop {} {\emph {\bibinfo {title} {Nonextensive
  Entropy---Interdisciplinary Applications}}}\ (\bibinfo  {publisher} {Oxford
  University Press},\ \bibinfo {address} {New York},\ \bibinfo {year}
  {2004})\BibitemShut {NoStop}%
\bibitem [{\citenamefont {{PHENIX Collaboration}}(2011)}]{Adare2011}%
  \BibitemOpen
  \bibfield  {author} {\bibinfo {author} {\bibnamefont {{PHENIX
  Collaboration}}},\ }\href@noop {} {\bibfield  {journal} {\bibinfo  {journal}
  {Phys. Rev. D}\ }\textbf {\bibinfo {volume} {83}},\ \bibinfo {pages} {052004}
  (\bibinfo {year} {2011})}\BibitemShut {NoStop}%
\bibitem [{\citenamefont {Afsar}\ and\ \citenamefont
  {Tirnakli}(2013)}]{Afsar2013}%
  \BibitemOpen
  \bibfield  {author} {\bibinfo {author} {\bibfnamefont {O.}~\bibnamefont
  {Afsar}}\ and\ \bibinfo {author} {\bibfnamefont {U.}~\bibnamefont
  {Tirnakli}},\ }\href@noop {} {\bibfield  {journal} {\bibinfo  {journal}
  {EPL}\ }\textbf {\bibinfo {volume} {101}},\ \bibinfo {pages} {20003}
  (\bibinfo {year} {2013})}\BibitemShut {NoStop}%
\bibitem [{\citenamefont {Lutz}\ and\ \citenamefont
  {Renzoni}(2013)}]{Lutz2013}%
  \BibitemOpen
  \bibfield  {author} {\bibinfo {author} {\bibfnamefont {E.}~\bibnamefont
  {Lutz}}\ and\ \bibinfo {author} {\bibfnamefont {F.}~\bibnamefont {Renzoni}},\
  }\href@noop {} {\bibfield  {journal} {\bibinfo  {journal} {Nat. Phys.}\
  }\textbf {\bibinfo {volume} {9}},\ \bibinfo {pages} {615} (\bibinfo {year}
  {2013})}\BibitemShut {NoStop}%
\bibitem [{\citenamefont {Vallianatos}(2013)}]{Vallianatos2013}%
  \BibitemOpen
  \bibfield  {author} {\bibinfo {author} {\bibfnamefont {F.}~\bibnamefont
  {Vallianatos}},\ }\href@noop {} {\bibfield  {journal} {\bibinfo  {journal}
  {EPL}\ }\textbf {\bibinfo {volume} {102}},\ \bibinfo {pages} {28006}
  (\bibinfo {year} {2013})}\BibitemShut {NoStop}%
\bibitem [{\citenamefont {Gontis}\ and\ \citenamefont
  {Kononovicius}(2014)}]{Gontis2014}%
  \BibitemOpen
  \bibfield  {author} {\bibinfo {author} {\bibfnamefont {V.}~\bibnamefont
  {Gontis}}\ and\ \bibinfo {author} {\bibfnamefont {A.}~\bibnamefont
  {Kononovicius}},\ }\href@noop {} {\bibfield  {journal} {\bibinfo  {journal}
  {PLoS ONE}\ }\textbf {\bibinfo {volume} {9}},\ \bibinfo {pages} {e102201}
  (\bibinfo {year} {2014})}\BibitemShut {NoStop}%
\bibitem [{\citenamefont {Tanaka}\ \emph {et~al.}(2015)\citenamefont {Tanaka},
  \citenamefont {Nishikawa}, \citenamefont {Kurths}, \citenamefont {Chen},\
  and\ \citenamefont {Kiss}}]{Tanaka2015}%
  \BibitemOpen
  \bibfield  {author} {\bibinfo {author} {\bibfnamefont {H.-A.}\ \bibnamefont
  {Tanaka}}, \bibinfo {author} {\bibfnamefont {I.}~\bibnamefont {Nishikawa}},
  \bibinfo {author} {\bibfnamefont {J.}~\bibnamefont {Kurths}}, \bibinfo
  {author} {\bibfnamefont {Y.}~\bibnamefont {Chen}}, \ and\ \bibinfo {author}
  {\bibfnamefont {I.~Z.}\ \bibnamefont {Kiss}},\ }\href@noop {} {\bibfield
  {journal} {\bibinfo  {journal} {EPL}\ }\textbf {\bibinfo {volume} {111}},\
  \bibinfo {pages} {50007} (\bibinfo {year} {2015})}\BibitemShut {NoStop}%
\bibitem [{\citenamefont {Jauregui}\ and\ \citenamefont
  {Tsallis}(2015)}]{Jauregui2015}%
  \BibitemOpen
  \bibfield  {author} {\bibinfo {author} {\bibfnamefont {M.}~\bibnamefont
  {Jauregui}}\ and\ \bibinfo {author} {\bibfnamefont {C.}~\bibnamefont
  {Tsallis}},\ }\href@noop {} {\bibfield  {journal} {\bibinfo  {journal} {J.
  Math. Phys.}\ }\textbf {\bibinfo {volume} {56}},\ \bibinfo {pages} {023303}
  (\bibinfo {year} {2015})}\BibitemShut {NoStop}%
\bibitem [{\citenamefont {Tsallis}(2009{\natexlab{b}})}]{Tsallis2009-2}%
  \BibitemOpen
  \bibfield  {author} {\bibinfo {author} {\bibfnamefont {C.}~\bibnamefont
  {Tsallis}},\ }\href@noop {} {\bibfield  {journal} {\bibinfo  {journal} {Braz.
  J. Phys.}\ }\textbf {\bibinfo {volume} {39}},\ \bibinfo {pages} {337}
  (\bibinfo {year} {2009}{\natexlab{b}})}\BibitemShut {NoStop}%
\bibitem [{\citenamefont {Hanel}\ and\ \citenamefont
  {Thurner}(2011)}]{Hanel2011-1}%
  \BibitemOpen
  \bibfield  {author} {\bibinfo {author} {\bibfnamefont {R.}~\bibnamefont
  {Hanel}}\ and\ \bibinfo {author} {\bibfnamefont {S.}~\bibnamefont
  {Thurner}},\ }\href@noop {} {\bibfield  {journal} {\bibinfo  {journal} {EPL}\
  }\textbf {\bibinfo {volume} {93}},\ \bibinfo {pages} {20006} (\bibinfo {year}
  {2011})}\BibitemShut {NoStop}%
\bibitem [{\citenamefont {Hanel}\ \emph {et~al.}(2011)\citenamefont {Hanel},
  \citenamefont {Thurner},\ and\ \citenamefont {Gell-Mann}}]{Hanel2011-2}%
  \BibitemOpen
  \bibfield  {author} {\bibinfo {author} {\bibfnamefont {R.}~\bibnamefont
  {Hanel}}, \bibinfo {author} {\bibfnamefont {S.}~\bibnamefont {Thurner}}, \
  and\ \bibinfo {author} {\bibfnamefont {M.}~\bibnamefont {Gell-Mann}},\
  }\href@noop {} {\bibfield  {journal} {\bibinfo  {journal} {PNAS}\ }\textbf
  {\bibinfo {volume} {108}},\ \bibinfo {pages} {6390} (\bibinfo {year}
  {2011})}\BibitemShut {NoStop}%
\bibitem [{\citenamefont {Ruiz}\ and\ \citenamefont
  {Tsallis}(2015)}]{Ruiz2015}%
  \BibitemOpen
  \bibfield  {author} {\bibinfo {author} {\bibfnamefont {G.}~\bibnamefont
  {Ruiz}}\ and\ \bibinfo {author} {\bibfnamefont {C.}~\bibnamefont {Tsallis}},\
  }\href@noop {} {\bibfield  {journal} {\bibinfo  {journal} {J. Math. Phys.}\
  }\textbf {\bibinfo {volume} {56}},\ \bibinfo {pages} {053301} (\bibinfo
  {year} {2015})}\BibitemShut {NoStop}%
\bibitem [{\citenamefont {Rodriguez}\ \emph {et~al.}(2008)\citenamefont
  {Rodriguez}, \citenamefont {Schwammle},\ and\ \citenamefont
  {Tsallis}}]{Rodriguez2008}%
  \BibitemOpen
  \bibfield  {author} {\bibinfo {author} {\bibfnamefont {A.}~\bibnamefont
  {Rodriguez}}, \bibinfo {author} {\bibfnamefont {V.}~\bibnamefont
  {Schwammle}}, \ and\ \bibinfo {author} {\bibfnamefont {C.}~\bibnamefont
  {Tsallis}},\ }\href@noop {} {\bibfield  {journal} {\bibinfo  {journal} {J.
  Stat. Mech.}\ }\textbf {\bibinfo {volume} {2008}},\ \bibinfo {pages} {P09006}
  (\bibinfo {year} {2008})}\BibitemShut {NoStop}%
\bibitem [{\citenamefont {Hanel}\ \emph {et~al.}(2009)\citenamefont {Hanel},
  \citenamefont {Thurner},\ and\ \citenamefont {Tsallis}}]{Hanel2009}%
  \BibitemOpen
  \bibfield  {author} {\bibinfo {author} {\bibfnamefont {R.}~\bibnamefont
  {Hanel}}, \bibinfo {author} {\bibfnamefont {S.}~\bibnamefont {Thurner}}, \
  and\ \bibinfo {author} {\bibfnamefont {C.}~\bibnamefont {Tsallis}},\
  }\href@noop {} {\bibfield  {journal} {\bibinfo  {journal} {Eur. Phys. J. B}\
  }\textbf {\bibinfo {volume} {72}},\ \bibinfo {pages} {263} (\bibinfo {year}
  {2009})}\BibitemShut {NoStop}%
\bibitem [{\citenamefont {Rodr{\'\i}guez}\ and\ \citenamefont
  {Tsallis}(2012)}]{Rodriguez2012}%
  \BibitemOpen
  \bibfield  {author} {\bibinfo {author} {\bibfnamefont {A.}~\bibnamefont
  {Rodr{\'\i}guez}}\ and\ \bibinfo {author} {\bibfnamefont {C.}~\bibnamefont
  {Tsallis}},\ }\href@noop {} {\bibfield  {journal} {\bibinfo  {journal} {J.
  Math. Phys.}\ }\textbf {\bibinfo {volume} {53}},\ \bibinfo {pages} {023302}
  (\bibinfo {year} {2012})}\BibitemShut {NoStop}%
\bibitem [{\citenamefont {Ruseckas}(2015)}]{Ruseckas2015}%
  \BibitemOpen
  \bibfield  {author} {\bibinfo {author} {\bibfnamefont {J.}~\bibnamefont
  {Ruseckas}},\ }\href@noop {} {\bibfield  {journal} {\bibinfo  {journal}
  {Phys. Lett. A}\ }\textbf {\bibinfo {volume} {379}},\ \bibinfo {pages} {654}
  (\bibinfo {year} {2015})}\BibitemShut {NoStop}%
\bibitem [{\citenamefont {Moyano}\ \emph {et~al.}(2006)\citenamefont {Moyano},
  \citenamefont {Tsallis},\ and\ \citenamefont {Gell-Mann}}]{Moyano2006}%
  \BibitemOpen
  \bibfield  {author} {\bibinfo {author} {\bibfnamefont {L.~G.}\ \bibnamefont
  {Moyano}}, \bibinfo {author} {\bibfnamefont {C.}~\bibnamefont {Tsallis}}, \
  and\ \bibinfo {author} {\bibfnamefont {M.}~\bibnamefont {Gell-Mann}},\
  }\href@noop {} {\bibfield  {journal} {\bibinfo  {journal} {Europhys. Lett.}\
  }\textbf {\bibinfo {volume} {73}},\ \bibinfo {pages} {813} (\bibinfo {year}
  {2006})}\BibitemShut {NoStop}%
\bibitem [{\citenamefont {Hilhorst}\ and\ \citenamefont
  {Schehr}(2007)}]{Hilhorst2007}%
  \BibitemOpen
  \bibfield  {author} {\bibinfo {author} {\bibfnamefont {H.~J.}\ \bibnamefont
  {Hilhorst}}\ and\ \bibinfo {author} {\bibfnamefont {G.}~\bibnamefont
  {Schehr}},\ }\href@noop {} {\bibfield  {journal} {\bibinfo  {journal} {J.
  Stat. Mech.}\ }\textbf {\bibinfo {volume} {2007}},\ \bibinfo {pages} {P06003}
  (\bibinfo {year} {2007})}\BibitemShut {NoStop}%
\bibitem [{\citenamefont {Borland}(1998{\natexlab{a}})}]{Borland1998}%
  \BibitemOpen
  \bibfield  {author} {\bibinfo {author} {\bibfnamefont {L.}~\bibnamefont
  {Borland}},\ }\href@noop {} {\bibfield  {journal} {\bibinfo  {journal} {Phys.
  Rev. E}\ }\textbf {\bibinfo {volume} {57}},\ \bibinfo {pages} {6634}
  (\bibinfo {year} {1998}{\natexlab{a}})}\BibitemShut {NoStop}%
\bibitem [{\citenamefont {Borland}(2002)}]{Borland2002}%
  \BibitemOpen
  \bibfield  {author} {\bibinfo {author} {\bibfnamefont {L.}~\bibnamefont
  {Borland}},\ }\href@noop {} {\bibfield  {journal} {\bibinfo  {journal} {Phys.
  Rev. Lett.}\ }\textbf {\bibinfo {volume} {89}},\ \bibinfo {pages} {098701}
  (\bibinfo {year} {2002})}\BibitemShut {NoStop}%
\bibitem [{\citenamefont {Anteneodo}\ and\ \citenamefont
  {Tsallis}(2003)}]{Anteneodo2003}%
  \BibitemOpen
  \bibfield  {author} {\bibinfo {author} {\bibfnamefont {C.}~\bibnamefont
  {Anteneodo}}\ and\ \bibinfo {author} {\bibfnamefont {C.}~\bibnamefont
  {Tsallis}},\ }\href@noop {} {\bibfield  {journal} {\bibinfo  {journal} {J.
  Math. Phys.}\ }\textbf {\bibinfo {volume} {72}},\ \bibinfo {pages} {5194}
  (\bibinfo {year} {2003})}\BibitemShut {NoStop}%
\bibitem [{\citenamefont {dos Santos}\ and\ \citenamefont
  {Tsallis}(2010)}]{Santos2010}%
  \BibitemOpen
  \bibfield  {author} {\bibinfo {author} {\bibfnamefont {B.~C.}\ \bibnamefont
  {dos Santos}}\ and\ \bibinfo {author} {\bibfnamefont {C.}~\bibnamefont
  {Tsallis}},\ }\href@noop {} {\bibfield  {journal} {\bibinfo  {journal} {Phys.
  Rev. E}\ }\textbf {\bibinfo {volume} {82}},\ \bibinfo {pages} {061119}
  (\bibinfo {year} {2010})}\BibitemShut {NoStop}%
\bibitem [{\citenamefont {Queiros}\ \emph {et~al.}(2007)\citenamefont
  {Queiros}, \citenamefont {Moyano}, \citenamefont {{de Souza}},\ and\
  \citenamefont {Tsallis}}]{Queiros2007}%
  \BibitemOpen
  \bibfield  {author} {\bibinfo {author} {\bibfnamefont {S.~M.~D.}\
  \bibnamefont {Queiros}}, \bibinfo {author} {\bibfnamefont {L.~G.}\
  \bibnamefont {Moyano}}, \bibinfo {author} {\bibfnamefont {J.}~\bibnamefont
  {{de Souza}}}, \ and\ \bibinfo {author} {\bibfnamefont {C.}~\bibnamefont
  {Tsallis}},\ }\href@noop {} {\bibfield  {journal} {\bibinfo  {journal} {Eur.
  Phys. J. B}\ }\textbf {\bibinfo {volume} {55}},\ \bibinfo {pages} {161}
  (\bibinfo {year} {2007})}\BibitemShut {NoStop}%
\bibitem [{\citenamefont {Beck}(2001)}]{Beck2001}%
  \BibitemOpen
  \bibfield  {author} {\bibinfo {author} {\bibfnamefont {C.}~\bibnamefont
  {Beck}},\ }\href@noop {} {\bibfield  {journal} {\bibinfo  {journal} {Phys.
  Rev. Lett.}\ }\textbf {\bibinfo {volume} {87}},\ \bibinfo {pages} {180601}
  (\bibinfo {year} {2001})}\BibitemShut {NoStop}%
\bibitem [{\citenamefont {Ruseckas}\ and\ \citenamefont
  {Kaulakys}(2011)}]{Ruseckas2011}%
  \BibitemOpen
  \bibfield  {author} {\bibinfo {author} {\bibfnamefont {J.}~\bibnamefont
  {Ruseckas}}\ and\ \bibinfo {author} {\bibfnamefont {B.}~\bibnamefont
  {Kaulakys}},\ }\href@noop {} {\bibfield  {journal} {\bibinfo  {journal}
  {Phys. Rev. E}\ }\textbf {\bibinfo {volume} {84}},\ \bibinfo {pages} {051125}
  (\bibinfo {year} {2011})}\BibitemShut {NoStop}%
\bibitem [{\citenamefont {Combe}\ \emph {et~al.}(2015)\citenamefont {Combe},
  \citenamefont {Richefeu}, \citenamefont {Stasiak},\ and\ \citenamefont
  {Atman}}]{Combe2015}%
  \BibitemOpen
  \bibfield  {author} {\bibinfo {author} {\bibfnamefont {G.}~\bibnamefont
  {Combe}}, \bibinfo {author} {\bibfnamefont {V.}~\bibnamefont {Richefeu}},
  \bibinfo {author} {\bibfnamefont {M.}~\bibnamefont {Stasiak}}, \ and\
  \bibinfo {author} {\bibfnamefont {A.~P.~F.}\ \bibnamefont {Atman}},\
  }\href@noop {} {\bibfield  {journal} {\bibinfo  {journal} {Phys. Rev. Lett.}\
  }\textbf {\bibinfo {volume} {115}},\ \bibinfo {pages} {238301} (\bibinfo
  {year} {2015})}\BibitemShut {NoStop}%
\bibitem [{\citenamefont {van Kampen}(2011)}]{Kampen2011}%
  \BibitemOpen
  \bibfield  {author} {\bibinfo {author} {\bibfnamefont {N.~G.}\ \bibnamefont
  {van Kampen}},\ }\href@noop {} {\emph {\bibinfo {title} {Stochastic Processes
  in Physics and Chemistry}}}\ (\bibinfo  {publisher} {Elsevier},\ \bibinfo
  {address} {Amsterdam},\ \bibinfo {year} {2011})\BibitemShut {NoStop}%
\bibitem [{\citenamefont {Gardiner}(2009)}]{Gardiner2009}%
  \BibitemOpen
  \bibfield  {author} {\bibinfo {author} {\bibfnamefont {C.}~\bibnamefont
  {Gardiner}},\ }\href@noop {} {\emph {\bibinfo {title} {Stochastic Methods: A
  Handbook for the Natural and Social Sciences}}}\ (\bibinfo  {publisher}
  {Springer},\ \bibinfo {address} {Berlin},\ \bibinfo {year}
  {2009})\BibitemShut {NoStop}%
\bibitem [{\citenamefont {Borland}(1998{\natexlab{b}})}]{Borland1998-2}%
  \BibitemOpen
  \bibfield  {author} {\bibinfo {author} {\bibfnamefont {L.}~\bibnamefont
  {Borland}},\ }\href@noop {} {\bibfield  {journal} {\bibinfo  {journal} {Phys.
  Lett. A}\ }\textbf {\bibinfo {volume} {245}},\ \bibinfo {pages} {67}
  (\bibinfo {year} {1998}{\natexlab{b}})}\BibitemShut {NoStop}%
\bibitem [{\citenamefont {Risken}\ and\ \citenamefont
  {Frank}(1996)}]{Risken1996}%
  \BibitemOpen
  \bibfield  {author} {\bibinfo {author} {\bibfnamefont {H.}~\bibnamefont
  {Risken}}\ and\ \bibinfo {author} {\bibfnamefont {T.}~\bibnamefont {Frank}},\
  }\href@noop {} {\emph {\bibinfo {title} {The Fokker-Planck Equation: Methods
  of Solution and Applications}}}\ (\bibinfo  {publisher} {Springer},\ \bibinfo
  {year} {1996})\BibitemShut {NoStop}%
\bibitem [{\citenamefont {Kaulakys}\ and\ \citenamefont
  {Ruseckas}(2004)}]{Kaulakys2004}%
  \BibitemOpen
  \bibfield  {author} {\bibinfo {author} {\bibfnamefont {B.}~\bibnamefont
  {Kaulakys}}\ and\ \bibinfo {author} {\bibfnamefont {J.}~\bibnamefont
  {Ruseckas}},\ }\href@noop {} {\bibfield  {journal} {\bibinfo  {journal}
  {Phys. Rev. E}\ }\textbf {\bibinfo {volume} {70}},\ \bibinfo {pages}
  {020101(R)} (\bibinfo {year} {2004})}\BibitemShut {NoStop}%
\bibitem [{\citenamefont {Kaulakys}\ \emph {et~al.}(2006)\citenamefont
  {Kaulakys}, \citenamefont {Ruseckas}, \citenamefont {Gontis},\ and\
  \citenamefont {Alaburda}}]{Kaulakys2006}%
  \BibitemOpen
  \bibfield  {author} {\bibinfo {author} {\bibfnamefont {B.}~\bibnamefont
  {Kaulakys}}, \bibinfo {author} {\bibfnamefont {J.}~\bibnamefont {Ruseckas}},
  \bibinfo {author} {\bibfnamefont {V.}~\bibnamefont {Gontis}}, \ and\ \bibinfo
  {author} {\bibfnamefont {M.}~\bibnamefont {Alaburda}},\ }\href@noop {}
  {\bibfield  {journal} {\bibinfo  {journal} {Physica A}\ }\textbf {\bibinfo
  {volume} {365}},\ \bibinfo {pages} {217} (\bibinfo {year}
  {2006})}\BibitemShut {NoStop}%
\bibitem [{\citenamefont {Ruseckas}\ and\ \citenamefont
  {Kaulakys}(2014)}]{Ruseckas2014}%
  \BibitemOpen
  \bibfield  {author} {\bibinfo {author} {\bibfnamefont {J.}~\bibnamefont
  {Ruseckas}}\ and\ \bibinfo {author} {\bibfnamefont {B.}~\bibnamefont
  {Kaulakys}},\ }\href@noop {} {\bibfield  {journal} {\bibinfo  {journal} {J.
  Stat. Mech.}\ }\textbf {\bibinfo {volume} {2014}},\ \bibinfo {pages} {P06005}
  (\bibinfo {year} {2014})}\BibitemShut {NoStop}%
\bibitem [{\citenamefont {Nobre}\ \emph {et~al.}(2012)\citenamefont {Nobre},
  \citenamefont {Souza},\ and\ \citenamefont {Curado}}]{Nobre2012}%
  \BibitemOpen
  \bibfield  {author} {\bibinfo {author} {\bibfnamefont {F.~D.}\ \bibnamefont
  {Nobre}}, \bibinfo {author} {\bibfnamefont {A.~M.~C.}\ \bibnamefont {Souza}},
  \ and\ \bibinfo {author} {\bibfnamefont {E.~M.~F.}\ \bibnamefont {Curado}},\
  }\href@noop {} {\bibfield  {journal} {\bibinfo  {journal} {Phys. Rev. E}\
  }\textbf {\bibinfo {volume} {86}},\ \bibinfo {pages} {061113} (\bibinfo
  {year} {2012})}\BibitemShut {NoStop}%
\bibitem [{\citenamefont {Nobre}\ \emph {et~al.}(2015)\citenamefont {Nobre},
  \citenamefont {Curado}, \citenamefont {Souza},\ and\ \citenamefont
  {Andrade}}]{Nobre2015}%
  \BibitemOpen
  \bibfield  {author} {\bibinfo {author} {\bibfnamefont {F.~D.}\ \bibnamefont
  {Nobre}}, \bibinfo {author} {\bibfnamefont {E.~M.~F.}\ \bibnamefont
  {Curado}}, \bibinfo {author} {\bibfnamefont {A.~M.~C.}\ \bibnamefont
  {Souza}}, \ and\ \bibinfo {author} {\bibfnamefont {R.~F.~S.}\ \bibnamefont
  {Andrade}},\ }\href@noop {} {\bibfield  {journal} {\bibinfo  {journal} {Phys.
  Rev. E}\ }\textbf {\bibinfo {volume} {91}},\ \bibinfo {pages} {022135}
  (\bibinfo {year} {2015})}\BibitemShut {NoStop}%
\end{thebibliography}

%

\end{document}